\documentclass[conference]{IEEEtran}
\usepackage{amsmath}
\usepackage{amssymb}
\usepackage[final]{graphicx}
\usepackage{floatflt}
\usepackage{color}
\usepackage{cite}
\usepackage{color}
\usepackage{wrapfig,subfigure}
\usepackage{hyperref}

\usepackage{amssymb,amsfonts,amsmath,amstext,amsthm,times}
\usepackage{pifont}
\usepackage{graphics}
\usepackage{epsfig,psfrag}
\usepackage{wrapfig}
\usepackage{times}
\usepackage{array}
\usepackage{epstopdf}
\usepackage{enumitem,kantlipsum}
\usepackage{float}
\usepackage{aliascnt}
\usepackage{caption}

\ifCLASSINFOpdf
\else
\fi
\DeclareMathAlphabet{\mathbfsl}{OT1}{ppl}{b}{it} 

\newcommand{\be}[1]{\begin{equation}\label{#1}}
\newcommand{\ee}{\end{equation}} 

\newcommand{\C}{{\Bbb C}} 

\renewcommand{\leq}{\leqslant}
 
\renewcommand{\geq}{\geqslant}

\newcommand{\Cref}[1]{Co\-ro\-lla\-ry\,\ref{#1}}

\def\x{\textbf{x}}

\def\y{\textbf{y}}
\newcommand{\stexp}{\mbox{$\mathbb{E}$}}   
\newcommand{\beq}{\begin{equation}}
\newcommand{\eeq}{\end{equation}}   
\newcommand{\bea}{\begin{eqnarray}}
\newcommand{\eea}{\end{eqnarray}}

\begin{document}
%
\title{A Bayesian Model for False Information Belief Impact, Optimal Design, and Fake News Containment}

\author{\IEEEauthorblockN{Amin Khajehnejad}
\IEEEauthorblockA{3Red Trading Group LLC., University of Michigan Ann Arbor\\
Email: aminkh@umich.edu, amin.khajehnejad@3redgroup.com}
\and
\IEEEauthorblockN{Shima Hajimirza}
\IEEEauthorblockA{Texas A\&M University \\
Email: shima.hm@tamu.edu}
}

%


\maketitle

\begin{abstract}
This  work is a technical approach to modeling false information nature, design, belief impact and containment in multi-agent networks. We present a Bayesian mathematical model for source information and viewer's belief, and how the former impacts the latter in a media (network) of broadcasters and viewers. Given the proposed model, we study how a particular information (true or false) can be optimally designed into a report, so that on average it conveys the most amount of the original intended information to the viewers of the network. Consequently, the model allows us to study susceptibility of a particular group of viewers to false information, as a function of statistical metrics of the their prior beliefs (e.g. bias, hesitation, open-mindedness, credibility assessment etc.). In addition, based on the same model we can study false information ``containment'' strategies imposed by network administrators. Specifically, we study a credibility assessment strategy, where every disseminated report must be within a certain distance of the truth. We study the trade-off between false and true information-belief convergence using this scheme which leads to ways for optimally deciding how truth sensitive an information dissemination network should operate. 
\end{abstract}


%
\IEEEpeerreviewmaketitle

\section{Introduction}
Adversarial spread of false information  based on exploitation of viewer's \underline{naivety}, \underline{gullibility} and \underline{bias} has been remarked numerously over social media and news networks, most notably in the midst of the most recent US presidential election and Brexit \cite{webD,election,brexit,post_truth,free_speech}. The notion is familiarly referred to as fake news. However, manipulative demonstration and spread of false information is not limited to news, and indisputably  exists in other domains such as advertising \cite{advertising1,advertising2,advertising3,advertising4}, financial markets (electronic trading) \cite{spoof1,spoof2}, cyber-security \cite{cybersecurity}, adversarial machine learning \cite{advML,secure_ml} and more, often referred to by terms such as misinformation or disinformation.  

\noindent  When it comes to studying false information spread, the goals are multifaceted and significantly vague. This is due to various reasons. Firstly, The nature of false information is  highly context-dependent and debatable. The notion of fake news for instance is not fully resolved and well-defined for many \cite{moral_reasoning,comedy0,comedy1}. The practice of fake news fabrication and broadcasting can take various forms, from  an individual creating and promoting fake social media posts, news websites etc. \cite{fn_creator}, to manipulative journalism of well-established news organizations doing so with the intention of appealing to a certain group of audience \cite{journalism_role}. False news creation and spread is also not necessarily bound to  malicious intention, and can be a mere reflection of erroneous, exaggerative or neglectful quoting of absorbed facts, rumors, panic epidemics and similar phenomena. Furthermore, false information spread can affect different systems in different ways, and consequently, administrators decide to handle it in various forms: identification, detection, prediction, regulation, education  or even ignorance are common mechanism that are usually adopted to combat false information \cite{policies}. Nevertheless, one needs to identify and study the following aspects of false information in their corresponding contexts:

\noindent {\bf1. Nature.}  What is false information, how is it originated/designed, and what distinguishes it from the truth?   
\\
\noindent {\bf2. Impact.} How does false information or fake news impact beliefs of the users/viewers of a system/network, and how does that in turn affect the dynamics of information flow and system operation?  
\\
\noindent {\bf 3. Detection.} What methodologies can be used to identify false information accurately and efficiently? 
\\
\noindent {\bf4. Spread (epidemic).} In an interactive network where information is propagated in a distributed way by users and outside feeds, what are the chances of epidemic false news spread as a function of network parameters?
\\
\noindent {\bf5. Containment and mitigation.} What mechanisms can prevent false information dissemination, or offset it with truth? At what costs can those strategies be implemented and, what are the expected outcomes and damages?
\\
\noindent Mathematical modeling of false information is far more challenging than semantic and social analysis, due to lack of basic fundamentals. Yet there are  a handful of recent works that have modeled various aspects of false information nature and spread  using technical methods. Most notably, the majority of works have investigated fraudulent accounts and credibility of news in online social networks such as twitter \cite{castillo2011information, gupta2012credibility, gupta2012evaluating, shariff2014user, thomas2013trafficking,yang2012analyzing}. A few other works have proposed models for disseminating \textit{good information} in social networks to mitigate the effect of \textit{misinformation}, thereby improving \textit{reliability} \cite{budak2011limiting,nguyen2012containment,zhang2016misinformation,ehsanfar2017incentivizing,user_action}. Misinformation and manipulation in financial limit order markets are probably studied far more in-depth, especially using game-theoretic approaches (see e.g. \cite{market_man_game,market_man_game_1,market_man_game_2,price_discovery}), due to much simpler and readily available numerical representations of information and beliefs. There are recent works that have  tried to adopt those approaches in the context of social networks. For example, the study of  \cite{ehsanfar2017incentivizing}  offers a game-theoretic approach to truth and fake news dissemination based on a
 mixed-strategy formulation of volunteer's dilemma in an adhoc  social network. However, those studies are still preliminary and immature compared to areas where numerical representation of information is available.   
\\



\noindent The current work is a technical (yet different) approach to false information design, impact and containment in multi-agent networks. We present a Bayesian mathematical model for information, viewer's belief, and how the former impacts the latter in a media (network) of broadcasters and viewers. Given the proposed model, we study how a particular information (true or false) can be optimally designed into a report, so that on average it conveys the most amount of the original intended information to the viewers of the network. The model then allows one to study susceptibility of a particular group of viewers to false information, as a function of statistical metrics of the their prior beliefs (e.g., bias, hesitation, open-mindedness, etc.). In addition, based on the same model we can study false information ``containment'' strategies imposed by network administrators. Specifically, we study a credibility assessment strategy, where every disseminated report must be within a certain distance of the truth. We study the trade-off between false and true information containment using this scheme which leads to ways for optimally deciding how truth sensitive an information dissemination network should operate.

\subsection{Proposed Model}
\begin{figure}[H]
\centering
\includegraphics[width=0.5\textwidth]{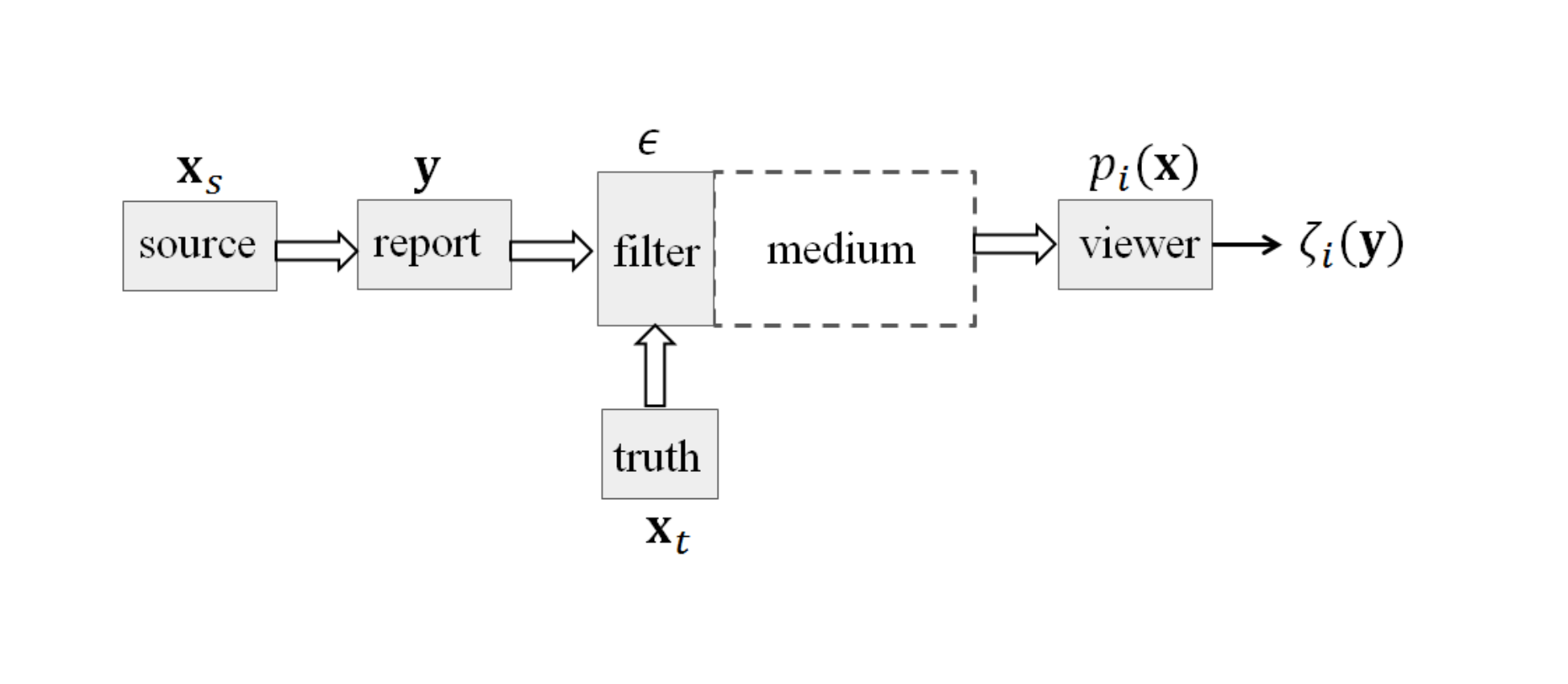}
\caption{\label{fig:network}Schematics of an information dissemination network with source, reporter, administrator and viewer.}
\end{figure}

We consider an information dissemination network (medium) with an information source, a transmitter (reporter), a network administrator and multiple viewers (audience). The schematics of such a network are illustrated in Figure \ref{fig:network}. 
\noindent The transmitter intends to disseminate in the network a piece of information $\x_s$ from the source. However, he can modify the information content and design a report $\y$, in order for the viewers to accept as much of the source information as possible. There is an authenticity filter in the network that verifies every input report by comparing it with the truth. The report is not permitted to be broadcast if it is too distant from the truth $\x_t$. Therefore, the reporter's optimal design must fall in an acceptable neighborhood of the truth which we denote by $\C(\x_t)$.  We assume that $\x_s,\x_t$ and $\y$ are all real vectors of length $n$. For the present, work we use the simple euclidean distance as the measure of deviation between facts: $d(\x_1,\x_2)=\|\x_1-\x_2\|$. Therefore, the authenticity filter has the following mathematical form:

\begin{equation}
\label{eq:filter}
\y \in \C(\x_t) = \{\x~|~\|\x - \x_t\| \leq \epsilon\}
\end{equation}

\noindent where $\epsilon$ is a positive constant. There are $N>>0$ viewers, and every viewer $i$ has a prior distribution $p_i(\x)$ about the truth. When a viewer receives a report $\y$, he/she forms a posterior distribution given the prior belief and the observation, and consequently adopts a new belief in a Bayesian manner, which we denote by $\zeta_i(\y)$. The details of the viewers belief impact is described in the next subsection. 
\noindent We make the assumption of the \textit{well-informed} reporter. This means that the reporter is fully aware of the truth $\x_s$, and also of the network administration policy $\epsilon$. 
However, the reporter is not aware of every individual's prior distribution. Instead, he only knows average statistics such as mean of first and seconds moments of the viewers' priors, as will be explained the next section. The reporter's goal is to convey the source's information to as many viewers as possible. For a single viewer $i$, information conveyance can be measured by the distance of the original information $\x_s$ and the adopted belief $\zeta_i(\y)$:
\beq 
\tau_{i,s}(\y) = \|\x_s - \zeta_i(\y)\| 
\eeq 

\noindent For simplicity of future reference, we call this quantity the ``reporter-viewer information convergence'', or in short ``convergence''. The smaller $\tau_{i,s}(\y)$, the more successful the reporter has been for impacting that particular viewer. However, the statistical notion of information throughput is obtained by averaging $\tau_{i,s}(\y)$ over all viewers:
\beq 
\tau_s(\y) = \stexp \|\x_s - \zeta_i(\y) \|
\label{eq:tau}
\eeq 
\noindent The reporter's goal is therefore to minimize average convergence $\tau_s(\y)$ for a valid report $\y$ which is permitted to pass through the network, i.e.

\bea
\underset{\y}{\mathrm{argmin}}{~\tau_s(\y)} \nonumber \\
\text{s.t.}~\y\in\C(\x_t) \label{eq:reporter_obj}
\eea

\subsection{Bayesian Viewer}
\noindent We model the viewer of the news (audience) as an agent $i$ that has a prior belief about the truth at a particular point in time. We denote the probability distribution of this prior with $p_i(\cdot)$. Upon receiving a piece of report $\y$, the viewer' instantaneous belief of the truth (his point of view) is the Maximum A Posteriori fact:
\beq 
\zeta_i(\y) = \underset{\x}{\mathrm{argmax}}{~p_i(\x|\y)},
\label{eq:zeta_initial}
\eeq 

\noindent where the posterior distribution of the true news is obtained from the  Bayes rule :
\beq 
p_i(\x|\y) \propto p_i(\x)p_i(\y|\x)
\label{eq:Bayes}
\eeq
\noindent The likelihood distribution $p_i(\y|\x)$ is a function of the \underline{uncertainty} and \underline{credibility} of the source of news in the viewer's mind. This is essentially an  estimation of the statistics of truth manipulation (in the mind of the viewer) that an information source performs on the facts before disseminating them. If in the viewer's mind, the source of the news is very noisy, exaggerative or inaccurate, then the likelihood distribution $p_i(\y|\x)$ is very wide and has a large variance. If there is a reporting bias to the journalism which is known to the viewer, then the distribution is nonzero mean. In summary, the viewer blends his prior opinion with what he/she observes, but takes into account the presumed credibility of the source (which could in all likelihood be false).     

\noindent In a perfectly Gaussian world, we can assume that the prior and likelihood are both $n$-dimensional Gaussian distributions: 
\bea p_i(\x)\sim\mathcal{N}(\mu_i,\Sigma_i),\nonumber\\  p_i(\x|\y)\sim\mathcal{N}(\y,\Sigma_{s,i}),
\label{eq:Gaussian}
\eea 
\noindent where $\Sigma_i$ and $\Sigma_{s,i}$ are covariance matrices of the viewer's prior and the conditional likelihood distribution, respectively. Intuitively, $\Sigma_i$ is a multi-dimensional measure of viewer's intrinsic hesitation about the state of truth at the time being, while $\Sigma_{s,i}$ measures suspicion of noisy/false reporting against the news source. The well-informed reporter assumption means that the reporter is not aware of individual viewer characteristics $\mu_i,\Sigma_{i},\Sigma_{s,i}$, but has perfect knowledge of the joint distribution of these quantities. Combining equations (\ref{eq:zeta_initial}),(\ref{eq:Bayes}),(\ref{eq:Gaussian}), we get:
\bea 
\log{p_i(\x|\y)} \propto-(\x-\mu_i)^T\Sigma_i^{-1}(\x-\mu_i) - (\x-\y)\Sigma_{s,i}^{-1}(\x-\y) 
\eea 

\noindent and by maximizing this, we get that the viewers post report belief becomes:

\beq
\boxed{\zeta_i(y)=A_i\y+B_i\mu_i},\\
\label{eq:zeta}
\eeq
\noindent where
\bea
A_i=\bigl(\Sigma_i^{-1} + \Sigma_{s,i}^{-1}\bigr)^{-1}\Sigma_{s,i}^{-1},\nonumber \\
B_i=\bigl(\Sigma_i^{-1} + \Sigma_{s,i}^{-1}\bigr)^{-1}\Sigma_{i}^{-1}
\eea

\noindent Therefore, in the world of Gaussian news and beliefs, a viewer adopts a linear combination of his prior belief and the viewed report as his/her new belief.

\subsection{Optimal Reporter}
Combining equations (\ref{eq:tau}),(\ref{eq:reporter_obj}) and (\ref{eq:zeta}) together, the optimal report $\y^*$ is revealed to be the solution of the following optimization: 
\begin{align}
&\underset{\y}{\min}~\stexp{\|A_i\y + B_i\mu_i -\x_s \|}, \nonumber \\
&\text{s.t.}~\|\y-\x_t\| \leq \epsilon. \label{eq:convex_opt}
\end{align}
\noindent (\ref{eq:convex_opt}) is a convex optimization. Therefore, using the method of Lagrange multipliers (see e.g., \cite{lagrange}), $\y^*$ is the minimizer of the following utility function for some nonnegative $\lambda\geq 0$:  
\bea 
L(\y,\lambda) = \stexp\|A_i\y + B_i\mu_i - \x_s\|^2 + \lambda \|\y-\x_t\|^2
\eea 
\noindent Taking derivatives with respect to $\y$, we get: 
\bea 
\frac{\partial{L(\y,\lambda)}}{\partial{\y}} = (\overline{A^2} + \lambda I)\y + \overline{AB\mu} - \overline{A}\x_s - \lambda \x_t
\label{eq:derivative}
\eea 
\noindent where:
\beq 
\overline{A^2} = \stexp{A_i^TA_i},~\overline{AB\mu} = \stexp{A_i^TB_i\mu},~ \overline{A}=\stexp{A_i}.
\eeq 
\noindent Finally, by setting (\ref{eq:derivative}) equal to zero, the optimal report is obtained as: 
\beq 
\boxed{
\y^* =  (\overline{A^2} + \lambda I)^{-1}\left(\lambda \x_t + \overline{A}\x_s -  \overline{AB\mu} \right),
}
\label{eq:opty}
\eeq 
\noindent where $\lambda$ is the smallest nonnegative constant that makes $\|\y^* - \x_t\| \leq \x_t$. The $\lambda$ in (\ref{eq:opty}) is a regularization factor that forces the report to be closer to reality. Note that the well-informed reporter assumption means that $\overline{A^2}$, $\overline{A}$  and $\overline{AB\mu}$ are computable for the reporter, hence so is the optimal report $\y^*$.  
\subsection{Ergodic Covariances}
For further simplicity of the analysis, we consider the case of ergodic covariances, i.e., $\Sigma_i=\Sigma$ and $\Sigma_{i,s}=\Sigma_s$ for all viewers, and only mean belief $\mu_i$ varies. Under this assumption, it is easy to show that 
\beq 
\y^* =  (A^2+ \lambda I)^{-1}\left(\lambda \x_t + A\x_s -  AB\overline{\mu} \right).
\eeq 

\noindent When $\lambda=0$, there is no authenticity requirement for the report, thus the optimal report can be designed solely based on the input source information and the viewers' belief statistics. Specifically: 
\beq 
\y^* =  A^{-1}\x_s -  A^{-1}B\overline{\mu}  = \x_s + \Sigma_s\Sigma^{-1}\left(\x_s - \overline{\mu}\right)   .
\eeq 

\noindent In other words, the optimal report is the linear sum of the news source and a  skewed version of the average news \textit{surprise} ($\x_s-\overline{\mu}$), which can also be interpreted as \textit{exaggeration}. The optimal magnitude and rotation of such exaggeration is determined by the relative ratio of the covariance matrices $\Sigma$ (hesitation) and $\Sigma_s$ (source reliability). The more reliable a source (smaller $\|\Sigma_s\|_*$) or the more open minded/less certain the viewers (larger $\|\Sigma\|_*$), more exaggeration is allowed. $\|X\|_*$ denotes the nuclear norm of a matrix $X$ which can be a measure of the magnitude of a covariance matrix. However, this also tells us that elements of the report (i.e. different coordinates of $\y$) can be exaggerated differently, according to the certainty distribution. It should be noted that even in the case of a perfectly truthful source $\x_s=\x_t$, the optimal report is exaggerative and not completely honest. A credibility assessment strategy or filtration policy should take that fact into consideration. In other words, enforcing truthful reporting will compensate the performance of an optimal reporter completely based on the truth if $\|\Sigma_s\Sigma^{-1}\left(\x_t - \overline{\mu}\right)\| >> \epsilon$. 
\noindent The optimal convergence metric $\tau_s(\y^*)$ for $\lambda=0$ in this case becomes:

\beq 
\tau_s(\y^*) = \stexp(\mu_i - \overline{\mu})^TB^2
(\mu_i - \overline{\mu}) 
\eeq 
\noindent which is independent of the particular piece of news $\x_s$, and only depends on the source credibility and viewers' belief statistics. 

\subsection{Authentication Policy}
Assuming that every reporter does the optimal 
action knowing the specified filter parameter $\epsilon$, it is up to the network administrator to set a value that promotes reporting based on truthful sources and limits the unauthentic ones. This requires definition of an appropriate truth dissemination utility function. For a given source $\x_s$, policy parameter $\epsilon$ and optimal report $\y^*$, we denote the optimal convergence with $c(\x_s, \epsilon)$, which can be obtained from the derivations of the previous section as:
\beq 
c(\x_s, \epsilon) = \underset{i}{\stexp} \|A\y^*_{\epsilon} + B\mu_i - \x_s \|
\eeq 

\noindent The criteria for $\epsilon$ in filter design is intuitive: average convergence for a source of false information must be significantly smaller than that of the authentic source. In addition, $\epsilon$ should not be not too small to limit the truthful optimal reporter. These  criteria can be represented by: 
\begin{align} 
&U_1(\epsilon)=\underset{\x_t,\x_s}\stexp\biggl(c(\x_s,\epsilon) - c(\x_t,\epsilon)~\biggl\lvert~\|\x_t - \x_s \| \geq d_{min}\biggr) \geq \delta,  \label{eq:util_1}\\
&U_2(\epsilon)=\frac{\underset{\x_t}\stexp{c(\x_t,\infty)}}{ \underset{\x_t}\stexp{c(\x_t,\epsilon)} } \geq \alpha,   \label{eq:util_2}
\end{align}
\noindent where $\delta$ and $\alpha$ are positive constants. The first equation states that the average convergence distance for false news information sources must be farther than that of the true information source by a minimum margin $\delta$. Note that the expectation is over the distribution of all information sources, and an information vector is considered false if it is at a distance $d_{min}$ from the truth. The second equation on the other hand guarantees that the network is not too restrictive for the optimal reporter of truth compared with a free network with no authentication policy. Combining these metrics, we can define a unified utility function, and choose the maximizer $\epsilon^*$:

\begin{align}
\epsilon^*&=\underset{\epsilon}{\mathrm{argmax}}~U(\epsilon) \nonumber \\
&= \underset{\x_t,\x_s}\stexp\biggl(c(\x_s,\epsilon) - c(\x_t,\epsilon)~\biggl\lvert~\|\x_t - \x_s \| \geq d_{min}\biggr) \nonumber \\
&+ \beta\frac{\underset{\x_t}\stexp{c(\x_t,\infty)}}{\underset{\x_t}\stexp{c(\x_t,\epsilon)}}  
\label{eq:overall_util}
\end{align}

\begin{figure}[h]
\centering
\includegraphics[width=0.5\textwidth]{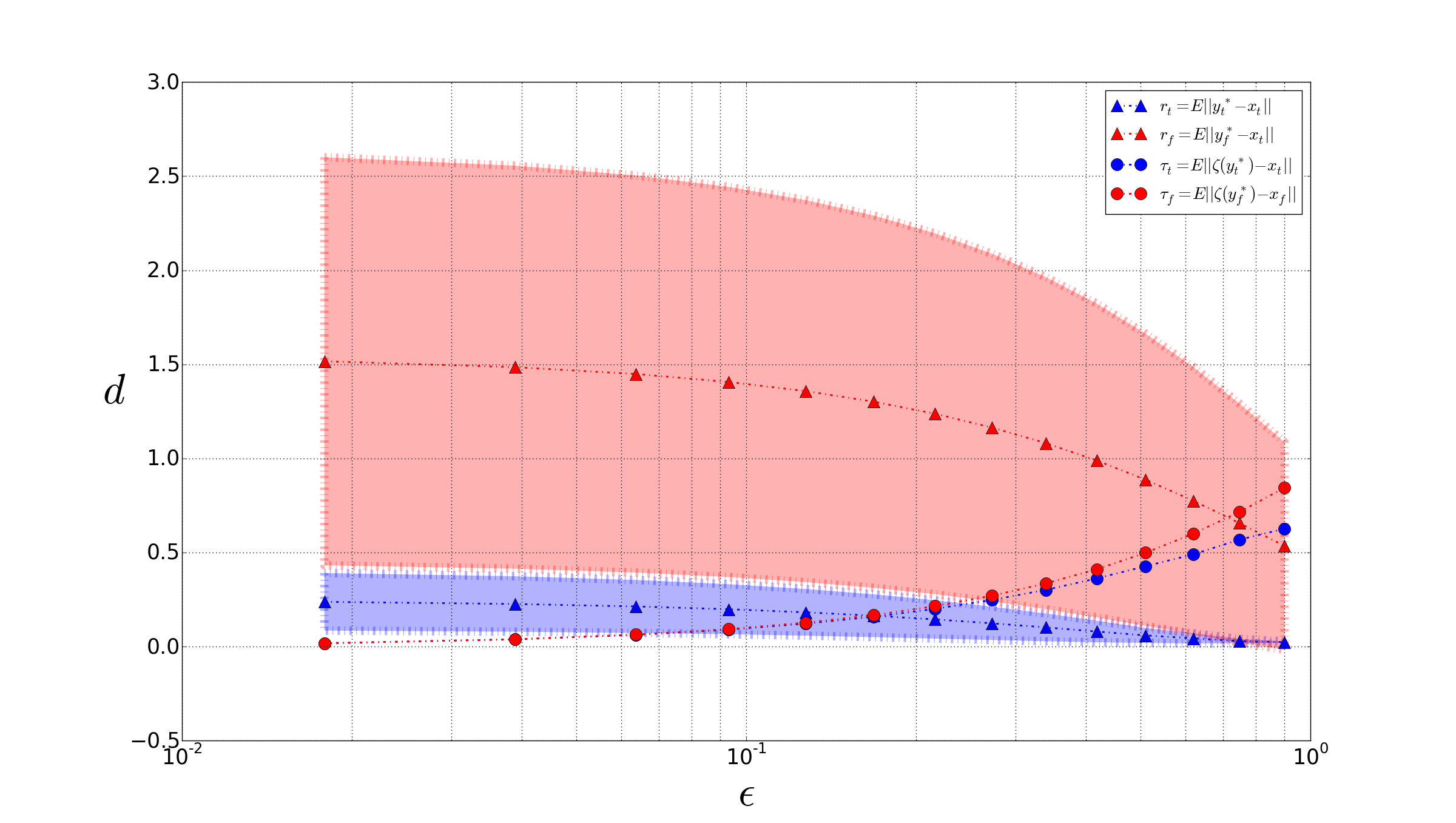}
\caption{Statistics of belief convergence  of optimal report for true information (blue) and random unauthentic information (red).} 
\label{fig:general_curve}
\end{figure}

\section{Numerical Evaluations}
We set up a basic numerical simulation to calculate and illustrate the convergence metrics and utility functions of true and unauthentic information sources based on the derivations of the previous section. We consider a simple case of $n=2$, with ergodic covariances $\Sigma,\Sigma_s$ given below in (\ref{eq:setup}). We assume that the true information $\x_t$ and source information $\x_s$ are both independently generated from a uniform distribution of vectors on the unit $2$-sphere, $S_2$. The average viewers' belief $\overline{\mu}$ is also drawn from the same distribution.

\begin{align} 
\label{eq:setup}
\Sigma = 
\begin{bmatrix}
    1       & 0 \\
    0       & 1
\end{bmatrix},~
\Sigma_s = &
\begin{bmatrix}
    0.5       & 0 \\
    0       & 0.5
\end{bmatrix},~
\mu_i\sim\mathcal{N}\left(\overline{\mu},\begin{bmatrix}
    0.1       & 0 \\
    0       & 0.1
\end{bmatrix}\right) \nonumber \\
&\overline{\mu},\x_t,\x_s\sim S^2,~
\end{align} 

\noindent In Figure \ref{fig:general_curve}, the  average convergence for true source $\stexp\|\zeta(\y^*_t)-\x_t\|$ is compared with the convergence of a random source $\stexp\|\zeta(\y^*_s)-\x_s\|$, as a function of network filter parameter $\epsilon$. The standard deviations of the same quantities are displayed as error areas. Since the mean belief is completely independent of the source and truth, this figure is a representative of a convergence trade-off in a network of indifferent audience. 

\begin{figure}[H]
\centering
\includegraphics[width=0.5\textwidth]{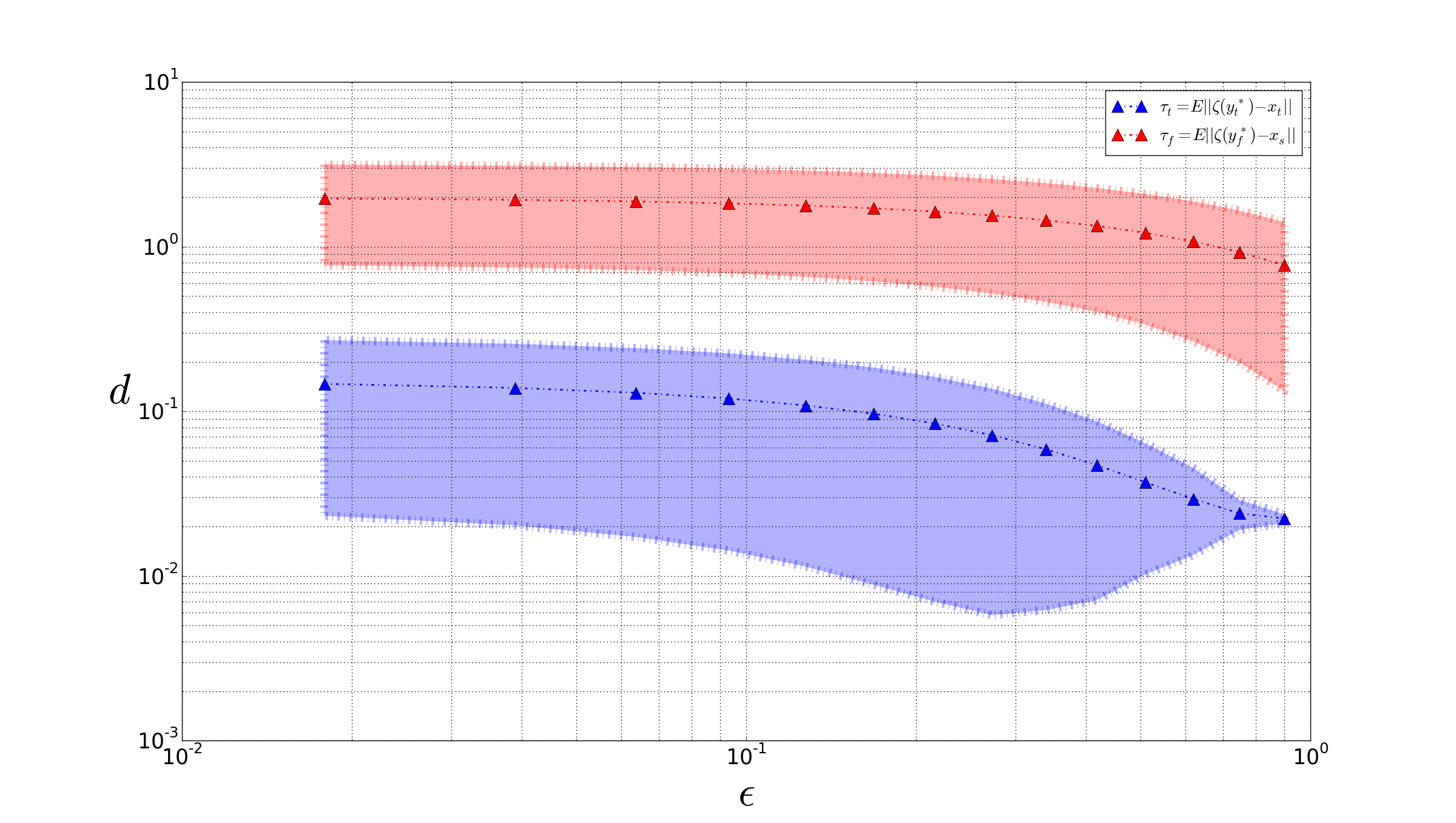}
\caption{Statistics of belief convergence  of optimal report for true information (blue) and random unauthentic information (red), for uneducated audience $\|\bar{\mu} - \x_s\| \leq \|\bar{\mu} - \x_t\|$.}
\label{fig:uneducated_curve}
\end{figure}

\noindent Figure \ref{fig:uneducated_curve} shows  convergence trade-off conditioned on the fact that average belief is closer to the source (false) information than the truth, i.e., $\|\bar{\mu} - \x_s\| \leq \|\bar{\mu} - \x_t\|$. We can think of this scenario as ``uneducated audience''. Note that for larger $\epsilon$, the false source convergence error area is below the average convergence of truth, meaning that the false information reporter has a significant chance of  conveying more information that the truth.         


\begin{figure}[H]
\centering
\includegraphics[width=0.5\textwidth]{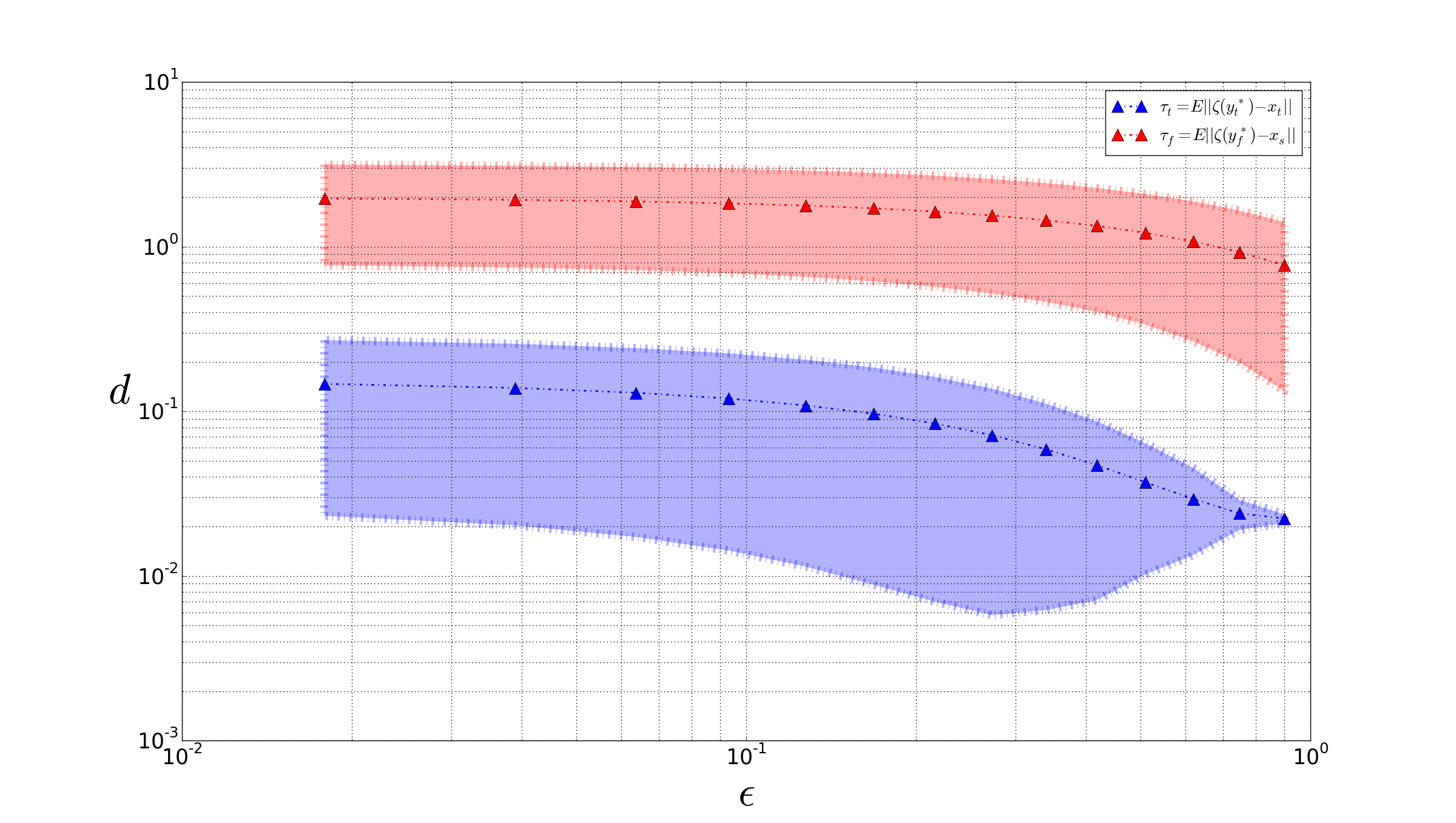}
\caption{ Statistics of belief convergence  of optimal report for true information (blue) and random unauthentic information (red), for educated audience $\|\bar{\mu} - \x_s\| \geq \|\bar{\mu} - \x_t\|$. }
\label{fig:educated_curve}.
\end{figure}

\noindent Figure \ref{fig:educated_curve} shows  convergence trade-off conditioned on the fact that average belief is farther to the source (false) information than the truth $\|\bar{\mu} - \x_s\| \geq \|\bar{\mu} - \x_t\|$.  

\begin{figure}[H]
\centering
\includegraphics[width=0.5\textwidth]{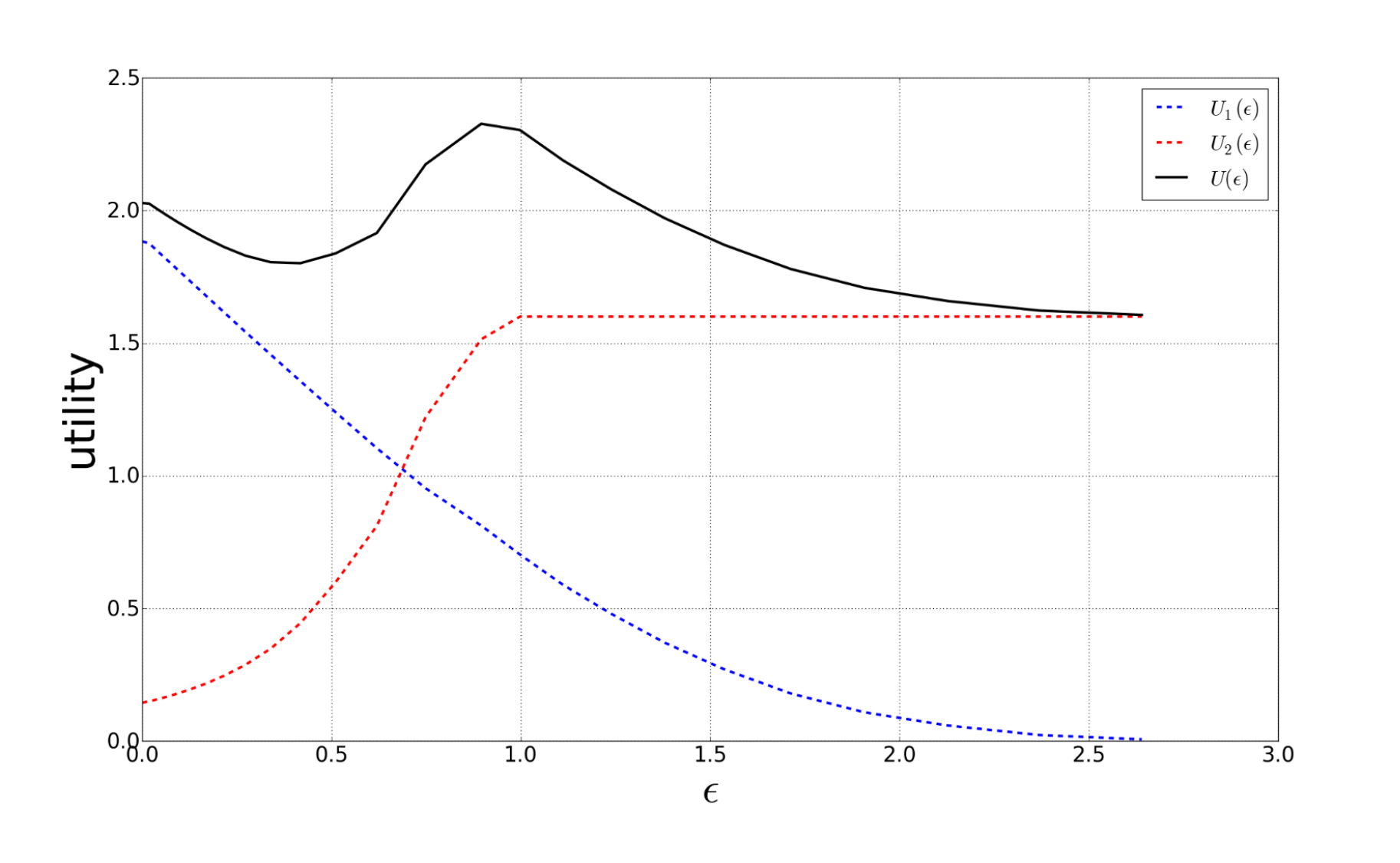}
\caption{Example of network utilities for $\beta=1.6,d_{min}=1.1$.} 
\label{fig:utilities}
\end{figure}

Finally, an example of network utility function for $\beta=1.6,d_{min}=1.1$ as a function of $\epsilon$ is shown in Figure \ref{fig:utilities}.

\section{Conclusion and Future Work}
A Bayesian model for information, viewer's belief and the impact of the former on the latter was proposed in this work. The model was studied from the point of an optimal reporter intending to convey the message of a particular source to a statistically large group of viewers, through a network that implements a basic authentication policy. The simplifying assumption of numerical information vectors and Gaussian priors, posteriors and belief impact was made. The reporter was also assumed to be well-informed, which means he has the knowledge of truth, network policy and joint distribution of viewers' belief statistics. Consequently, we derived the formulation of the optimal reporter, and showed that it is a linear combination of a portion of truth, source information and exaggeration. In addition, knowing this optimal strategy, we studied the optimal choice of the network authentication policy parameter. If the policy is too restrictive, it hurdles the performance of a well-intended reporter that only uses exaggerative techniques to convey the truth to an uneducated or biased audience. However, too loose of a policy grants a large capacity for false information dissemination to false or malicious sources. We proposed a computable utility function that takes these factors into account, and can be optimized for the best choice of authentication policy. 

\noindent Given that this study is the first of its own kind, the proposed model is expected to be significantly more progressed in the future work, in order to account for more practicalities. First and foremost, our primary goal for a follow-up work would be to validate the model with real-world data in various contexts (social media, news websites and electronic trading platforms).  Secondly, we intend to incorporate less perfect assumptions such partially informed reporters, and selective or random authentication policies, and provide more analytical and concrete derivations than those of the current work. 

\noindent Another limitation of the current model is that it is very static. In reality the source credibility parameters $\Sigma_{s,i}$ is a function of the long term reporting of a source, and varies through the time. Furthermore,  viewers have ways of cross-validating and comparing information coming from different sources. Other model parameters are also expected to be dynamic. 

\noindent The present model is also passive, which means the viewers do not spread the information, and are only receivers of it. A very interesting study would be to adopt the current model and study dynamics of belief evolution and propagation in a network where viewers act as two-way agents passing and receiving information.

\noindent Finally, the role of education can be explained and studied with the current model to some extent.  Education can be considered as a meta-function that skews viewers beliefs towards the truth over time. Given the importance of education in the age of social media and abundance of information, we believe that it is essential to provide modeling tools for studying its role in prevention of misinformation.






%




\end{document}